\documentstyle[11pt,epsf]{article}

\advance \topmargin by -\headheight
\advance \topmargin by -\headsep
\evensidemargin \oddsidemargin
\begin{document}    

\renewcommand{\thefootnote}{\fnsymbol{footnote}}


\def\footnoterule{\kern-3pt \hrule width \hsize \kern2.5pt}
\pagestyle{empty}

\begin{center}
{\Large\bf Evidence for String Substructure}

\vskip 1cm
\large{Oren Bergman}\footnote{E-mail  address: oren@phys.ufl.edu}
\vskip 0.5cm
{\it Institute for Fundamental Theory\\
Department of Physics, University of Florida, Gainesville,
FL, 32611, USA }

\end{center}

\vspace{1.2cm}
\begin{center}
{\bf ABSTRACT}
\end{center}

\noindent We argue that the behavior of string theory at
high temperature and high longitudinal boosts, 
combined the emergence of $p$-branes as necessary ingredients
in various string dualities, point to a possible reformulation
of strings, as well as $p$-branes, as composites of bits. We 
review
the string-bit models, and suggest generalizations to 
incorporate $p$-branes.

\vfill
\newpage

\def\balpha{\mbox{\boldmath$\alpha$}}
\def\bgamma{\hbox{\twelvembf\char\number 13}}
\def\bsigma{\hbox{\twelvembf\char\number 27}}
\def\bepsilon{\hbox{\twelvembf\char\number 15}}
\def\sgone{{\cal S}_1{\cal G}}
\def\sgtwo{{\cal S}_2{\cal G}}
\def\ob{\overline}
\def\bx{{\bf x}}
\def\by{{\bf y}}
\def\bz{{\bf z}}
\def\D{{\cal D}}
\def\R{{\cal R}}
\def\Q{{\cal Q}}
\def\G{{\cal G}}
\def\O{{\cal O}}
\def\N{{\cal N}}
\def\Nlarge{N_c\rightarrow\infty}
\def\Tr{{\rm Tr}}
\newcommand{\ket}[1]{|#1\rangle}
\newcommand{\bra}[1]{\langle#1|}
\newcommand{\firstket}[1]{|#1)}
\newcommand{\firstbra}[1]{(#1|}
\renewcommand{\thesection}{\arabic{section}}
\renewcommand{\thefootnote}{\arabic{footnote}}
\setcounter{footnote}{0}

\pagestyle{plain}
\pagenumbering{arabic}

\section{Introduction}
It has long been suspected that string theory
contains far fewer degrees of freedom than implied by the
world-sheet formulation, and it has been
suggested that a proper formulation should make this 
reduction manifest \cite{atickw}. A similar 
conclusion was reached
in \cite{susskindbh} in trying to explain the microscopic
origin of black hole entropy using string theory.
The area law seems to imply that 
the number of physical degrees of freedom in string
theory has to be severely reduced at trans-Planckian
energies.

Recent progress in the understanding of various 
dualities between string theories \cite{hullt,wittenduality}
has raised the question of whether there is 
a more fundamental formulation that encompasses 
all the known string theories. 
Eleven dimensional M-theory \cite{schwarzm,horavaw}
is frequently cited as a 
candidate for 
such a formulation, even though its precise
definition is not yet available. 
The type IIA and $E_8\times E_8$ string theories
result from distinct compactifications of the eleventh
dimension. The type IIB and $SO(32)$ string theories
are not directly reproduced by M-theory, but 
their compactifications are.  
If, as conjectured,
M-theory is a theory of membranes,
it raises the question of whether
string theory is really a theory of strings,
or whether strings, membranes and other
extended objects are equal members in a 
``$p$-brane democracy'' \cite{townsenddemocracy}.

Unrelated developments in string theory
\cite{thornmosc,thornrpa,bergmantbits,bergmantuniv} based
on the light-cone formulation have shown that it is 
possible to construct composite string-bit models for strings.
String-bit models
are non-relativistic (Galilean invariant) matrix field theories
of point particles (string-bits)
in $D-2$ space dimensions and one time dimension. They allow
for the formation of long closed chains, which in the 
continuum limit behave precisely as relativistic
strings. The $D-2$ space dimensions of the string-bits become
the transverse coordinates of the light-cone string, the time
dimension becomes $x^+$, and the total mass of the chain
becomes the longitudinal momentum of the string $p^+$.
Poincar\'e invariance is recovered
in the continuum limit at critical dimensions ($D=26$ for 
bosonic
string, $D=10$ for superstring).
These models automatically
incorporate chain splitting and joining processes that
mimic the closed string interactions.

At face value string-bit models are nothing more 
than discretizations
of string theories. They regularize 
string theory by providing an infrared cutoff
on the longitudinal momentum $p^+$, given by the mass of a bit.
Alternatively, one can take the point of view
that such a model is {\bf fundamental}, and that string
theory, as well as $D$-dimensional Poincar\'e invariance,
are low energy effective features.
We will argue that recent results in the study of
string dualities, together with older results concerning
the high temperature and high longitudinal boost
 behavior of string
theory lend some support to this point of view.

In section 2 we review the 
high temperature behavior of string theory
and its implication on the number of degrees of 
freedom in string theory. 
In section 3 we review
the ideas of Susskind on the behavior
of string theory at high longitudinal boosts, 
and how it is related to black
hole entropy. In section 4 
we review Townsend's idea of ``$p$-brane democracy'', that
arises from string dualities, and argue that just as in
``nuclear democracy'' the resolution lies in compositeness.
In section 5 we review string-bit models
in both the bosonic and
supersymmetric cases, and address the issues of the 
previous sections. 
In section 6 we suggest a generalization of the
string-bit models to incorporate higher dimensional
extended objects ($p$-branes).
Finally, section 7 is devoted to
a discussion of the successes and limitations of our 
particular 
composite models.

\section{High Temperature}
\renewcommand{\theequation}{\arabic{section}.\arabic{equation}}
The idea that there are degrees of freedom more fundamental
than strings in string theory was first suggested in 
\cite{atickw},
by studying the high-temperature behavior of string theory.
It has been well known from the early days of string theory
that the string partition function
diverges at temperatures $T>T_H$, where $T_H$ is
the Hagedorn temperature given by
\begin{equation}
T_H = {1\over 4\pi\sqrt{\alpha^\prime}}\; .
\end{equation}
Rather than being a limiting temperature it is believed
that $T_H$ is 
associated with a phase transition in string theory.

The appropriate target space for a finite temperature
ensemble of strings is $R^{D-1}\times S^1$, with
the radius of the $S^1$ given by $1/T$. 
This target space supports two winding modes which become tachyonic
for $T>T_H$. Precisely at $T=T_H$ these states are massless,
signaling a phase transition. 
The classical free energy above $T_H$ was computed in
\cite{atickw}, and is of the form
\begin{equation}
{F\over VT} \sim {1\over g^2T}\; .
\label{genus0}
\end{equation}
Since a genus-$k$ contribution to the free
energy is of order $g^{2(k-1)}$, the above must be a genus-$0$
contribution.
But since a genus-$0$ Riemann surface is simply connected,
i.e. has no nontrivial cycles, it cannot wind around $S^1$
to give a nontrivial temperature dependence.
The nontrivial result (\ref{genus0}) then signals a breakdown of 
the 
Riemann surface picture in the high-temperature phase.
The ``genus-$0$'' Riemann surface is somehow becoming
topologically nontrivial.

The large $N_c$ limit of QCD was invoked as support for 
this argument \cite{thornhotqcd}. QCD is believed to undergo a deconfining
phase transition at some temperature $T_{\rm dec}$. Below
$T_{\rm dec}$ the large $N_c$ limit of QCD is described in
terms of effective Riemann surfaces, and the free energy
is computed by counting glueball and meson states. Since
their spectrum is independent of $N_c$ for large $N_c$, the free
energy below $T_{\rm dec}$ is $\O(1)$. Above $T_{\rm dec}$
we know that the fundamental degrees of freedom are gluons
(and quarks), and therefore the free energy in the large $N_c$
limit is $\O(N_c^2)$.
Since the genus-$k$ contribution in the Riemann surface 
picture would be $(N_c^2)^{1-k}$, we see that below $T_{\rm dec}$
there are only genus$\geq 1$ contributions, whereas above
$T_{\rm dec}$ there is a genus-$0$ contribution as well. 
This signals a breakdown of the Riemann surface description
above $T_{\rm dec}$, which is precisely what happens.
Feynman diagrams involving quarks and gluons are Riemann
surfaces with ``holes''. Confinement below $T_{\rm dec}$
fills in the holes and gives smooth Riemann surfaces.
With the presence of holes, the ``genus-$0$ Riemann surface''
is no longer simply connected, and can contribute to the
free energy. 
In this case we know what the Riemann surface 
picture has to be replaced with at high temperature, namely 
a theory of quarks and gluons which are the composites
of the low temperature degrees of freedom. 
We will argue that 
it is quite plausible that the same must be done for string
theory.

The computation of the genus$\geq 1$ contributions to the
free energy in string theory gives an even more suggestive 
result than the genus-$0$ contribution. It was 
found that the genus-$k$ contribution is of the form
\begin{equation}
{F_k\over VT} \sim T(g^2T^2)^{k-1}\; .
\end{equation}
It was then argued that in order to get a sensible
high temperature limit, the effective coupling constant 
must be given by
\begin{equation}
g^2_{\rm eff}=g^2 T^2 \; ,
\end{equation}
which implies that the temperature dependence of the free
energy at high temperature must be
\begin{equation}
{F\over VT} \sim T \; .
\label{stfree}
\end{equation}
This is to be contrasted with the temperature dependence 
of the free energy of generic relativistic quantum field 
theories, given by
\begin{equation}
{F\over VT} \sim T^{D-1}\; ,
\end{equation}
indicating that there are far fewer gauge invariant 
degrees of freedom in string theory than in any relativistic
field theory in $D>2$ dimensions, 
let alone in the world-sheet formulation
of string theory.

The explanation given for this severe reduction in the number
of degrees of freedom was that at distances shorter than 
$\sqrt{\alpha^\prime}$ string theory is no longer properly
described in terms of strings or continuous world-sheets.
In fact the idea that $\sqrt{\alpha^\prime}$ serves as a 
minimum distance in string theory is also suggested by 
T-duality,
and by results on high-energy fixed-angle scattering
\cite{grossm}.
Whatever proper formulation of string theory describes
the breakdown of the Riemann surface picture at this scale,
should therefore also explain the scarcity of gauge invariant
degrees of freedom. This formulation clearly cannot be a
relativistic field theory, unless $D=2$.
A possibility not considered in \cite{atickw} is that the 
proper
formulation might be a non-relativistic (Galilean) field
theory. It is well known that a gas of non-relativistic 
non-interacting particles 
exhibits a free energy of the form
\begin{equation}
{F\over VT} \sim T^{(D-1)/2} \; ,
\label{nrfree}
\end{equation}
so that in two space dimensions ($D=3$) this yields the 
same behavior as
(\ref{stfree}). This is precisely the sort of system that will
be considered in section 5.

\section{High Longitudinal Boost}
\setcounter{equation}{0}
Results cited in the previous section imply the breakdown
of the world sheet (perturbative) formulation of string theory
at a length scale $\sqrt{\alpha^\prime}$, and suggest
that a proper formulation will contain far fewer degrees
of freedom. 
Since these results rely on perturbative techniques, the precise
length scale can be quite different.
In fact an indirect argument made by Susskind \cite{susskindbh}, 
in trying to reconcile
the behavior of strings at high longitudinal boosts with
black hole entropy, suggests that that the dramatic reduction
in the number of degrees of freedom is associated with a much 
shorter length scale given by $g\sqrt{\alpha^\prime}$, 
where $g$ is the closed 
string coupling constant.

The argument is based on one of the oldest known
facts about string theory, namely that the physical
size of a relativistic string is infinite.
This is most naturally seen in the light-cone gauge formulation,
where the mean square transverse separation and mean square
longitudinal separation are found to diverge :
\begin{eqnarray}
R_\perp^2 \equiv \langle |x_\perp(\sigma)-x_\perp(0)|^2\rangle 
 &\sim & \sum_{n=1}^\infty {1\over n}\longrightarrow\infty\nonumber \\
R_l^2 \equiv \langle | x^-(\sigma)-x^-(0)|^2\rangle
 &\sim & \sum_{n=1}^\infty n \longrightarrow\infty \; .
\label{sizes}
\end{eqnarray}
Realistically speaking however, any experiment
designed to
measure these quantities will involve a finite resolution
time $\epsilon$, implying that the highest measurable
frequency is $1/\epsilon$.
This provides an upper cutoff on the above sums of 
$n_{\rm max} = p^+/\epsilon$ in string units, where $p^+$ is the 
longitudinal
momentum of the string. The measured sizes will then be of the form
\begin{equation}
R_\perp^2\sim\ln{p^+\over\epsilon} \;\;\; , \;\;\;
R_l^2\sim\left({p^+\over\epsilon}\right)^2\; .
\end{equation}
The above describes a peculiar growth pattern of string with 
longitudinal momentum, and thus with energy.
As the longitudinal momentum is increased,
when the string falls towards a black hole for example,
the string
gets longer and grows more dense, until it becomes space 
filling
at infinite $p^+$ (see Fig.(1)). 

\begin{figure*}[htb]
\epsfxsize=4.5in
\centerline{\epsffile{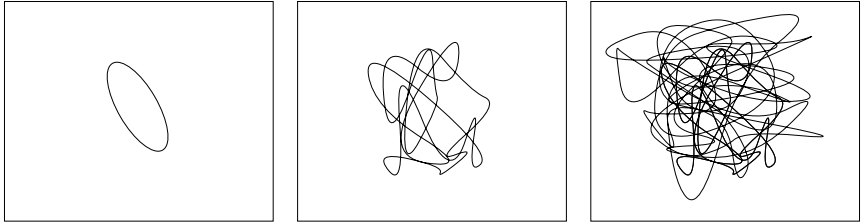}}
\caption{String growth with $p^+$}
\end{figure*}

This conclusion assumes that 
the perturbative formulation used to derive (\ref{sizes})
holds to arbitrarily high $p^+$, but it doesn't.
Eventually the transverse density of string, i.e. 
$R_l/R_\perp^{D-2}$, 
will exceed
$1/g^2$ (in string units, i.e. $\alpha^\prime =1$), at which point 
interactions 
will contribute at $\O(1)$ 
and perturbation theory will break down. So at best we can only 
trust this growth pattern up to this density.

For $D=4$ the density $1/g^2$ in string units is the the
inverse Planck area.
Susskind has suggested the possibility that 
the growth of string is simply 
cut off at the Planck scale \cite{susskindbh}.
The result is that the density grows 
until it reaches $\sim M_P^2$,
and then remains constant as the area of the transverse region
occupied by the string
grows linearly with $p^+$,
i.e.
\begin{equation}
R_\perp^2\sim {p^+\over\epsilon} \; .
\end{equation}
The motivation for this seemingly magical behavior of string
at the Planck scale comes from a piece of non-perturbative
information, namely the 
Beckenstein-Hawking formula for the entropy of a black hole,
\begin{equation}
S_{\rm BH} = {A\over 4G\hbar}\; .
\label{beck}
\end{equation}
To an outside observer this entropy should arise from
a counting of microstates associated with the stretched
horizon of the black hole \cite{susskindbh}, which
consists of infalling thermalized strings.
If the string density increases to the Planck density
and then remains fixed, it would imply that the black
hole entropy 
is proportional to
its area in Planck units as in (\ref{beck}).

Underlying this proposal is 't~Hooft's holographic principle 
\cite{thoofthologram}.
The Beckenstein-Hawking relation (\ref{beck}) puts an
upper
bound on the amount of information (entropy) that can be
fit into any compact 3-dimensional region of space, whether or
not there is a black hole there. Unlike our usual intuition
about entropy being an extensive quantity, the entropy is
bound by the area of the boundary of the region. 
In fact, 't~Hooft
has argued that the information contained in any 
compact region
of 3-dimensional space can be encoded on its 2-dimensional
boundary,
at a maximum density of one bit of information
per unit Planck area.
The 3-dimensional world is then merely a hologram,
since the true degrees of freedom come from 
two dimensions.
Perturbative light-cone string theory offers a glimpse 
of such a dimensional
reduction by eliminating all but the zero mode of the 
longitudinal
direction $x^-$. Susskind argues that non-perturbative 
effects should complete this picture by explaining 
the maximum density of information.

Discretization of the light-cone string eliminates even
the zero mode of $x^-$, by identifying its conjugate momentum
$p^+$ with the number of discrete bits in the string.
The string is replaced by a discrete chain of bits,
with a quantum-mechanical Hamiltonian whose continuum
limit is the string's $p^-$. If taken seriously, the 
discrete chain should arise as a stable bound state
of {\bf fundamental} microscopic constituents in a 
non-relativistic field theory in $2+1$ dimensions.
Thorn has devised string-bit models \cite{thornmosc} that 
achieve
precisely this, by utilizing 't Hoofts large $N_c$ limit.
The microscopic bit interactions would then have to explain
the origin of the maximum density, and the change in the 
growth pattern of the string.

\section{Duality and $p$-Brane Democracy}
\setcounter{equation}{0}
The past year and a half has produced a tremendous surge
of evidence for various dualities in string theory 
\cite{hullt,wittenduality}. An important
consequence of these new relations is
the understanding of the role played by
solitonic $p$-branes \cite{duffpbrane} in providing 
the non-perturbative BPS states required by duality. 
As BPS states these $p$-branes
preserve half the supersymmetry, and must therefore
carry $p$-form charges that enter the $D=10$ superalgebra
as central terms.
Upon compactification to four dimensions
these $p$-form central charges,
together with the internal momenta (Kaluza-Klein
and string winding modes), 
provide all the possible central charges of the 
$D=4$ superalgebra. The latter are of a 
perturbative origin, whereas the former are inherently
non-perturbative. However, from the point of view of the 
$D=4$ superalgebra there is no distinction.

The form of the superalgebra suggests that BPS-saturated
$p$-branes are possible, but it is the various conjectured
string dualities that make them necessary.
These dualities generically map BPS states to other
BPS states, and frequently map perturbative states
to non-perturbative ones. 
The non-perturbative BPS states necessarily  
correspond
to various wrapping modes of $p$-branes around $p$-cycles of the 
internal manifold.

For example in the conjectured duality 
of the
heterotic string on $T^4$ and
the type IIA string on $K3$ \cite{hullt,wittenduality},
the perturbative charged BPS states of the heterotic theory
map into non-perturbative R-R charged BPS states of the 
type IIA theory. 
Such states are known to exist as $p$-brane
solitons in $D=10$ type IIA supergravity,
with $p=0,2,4,6$, but have more recently been
understood from a stringy origin as D-branes 
\cite{polchinskidbrane}. 
Upon compactification on $K3$ the required states are given by
the $0$-brane, $2$-brane wrapped around any of the $22$ 
$2$-cycles of
$K3$, and $4$-brane wrapped around $K3$.

Similarly, the conjectured U-duality of type II $D=4$ string
theory \cite{hullt} maps perturbative BPS states
to non-perturbative ones in the same theory, and so
requires the wrapping modes of the $D=10$ $p$-branes as well.
Townsend has used this to argue that a non-perturbative 
formulation of string theory 
should exhibit a complete ``$p$-brane democracy''
\cite{townsenddemocracy}, in
which all $p$-branes enter as fundamental objects
\footnote{Townsend has made a distinction between
electrically charged $p$-branes and magnetically charged ones,
the former being more fundamental than the latter.}.
The string (1-brane) is singled out only because it
possesses a {\bf perturbative} formulation.

A possible setting in which at least the $D=10$
type IIA electric $p$-branes appear democratically is $D=11$
M-theory, whose low energy limit is $D=11$
supergravity.
It has been conjectured that M-theory is in fact 
a supermembrane theory \cite{townsendmembrane}. If so,
the type IIA $p=0,1,2$ branes all arise perturbatively from 
that membrane. The $p=4,5,6$ branes are then their
magnetic duals, and are therefore all non-perturbative.
This scenario is not quite satisfactory since there is still
a distinction between electric and magnetic $p$-branes,
and it does not address the type IIB
$p$-branes. More importantly, a successful quantization of the 
supermembrane is not yet available.

An amusing analogy can be drawn with the idea
of ``nuclear democracy'', advocated in the 60's by
Chew \cite{chewdemocracy}, and the emergence of the quark picture.
Nuclear democracy is the idea that all hadrons should be 
thought of as equally fundamental, even though 
particular formulations make some fundamental and others
not.
In the sigma
model, for example, mesons are fundamental and baryons appear 
as 
solitons. 
The ``Eightfold Way'' \cite{eightfold} showed that the 
baryons and mesons fit nicely into multiplets of a 
symmetry group $SU(3)$. This suggested a phenomenological
(but not necessarily dynamical) description of hadrons
as composed of quarks transforming in a smaller representation
of the symmetry group. Chew called this ``nuclear aristocracy'',
instead of all hadrons being equally fundamental they are 
all equally non-fundamental. This phenomenological
picture was subsequently given a dynamical setting with the 
formulation of QCD.

Duality is an idea similar in spirit to the eightfold way,
in that it groups BPS states
into ``multiplets''.
In particular it groups fundamental string states with
wrapping modes of $p$-brane solitons. In following this 
analogy through
we suggest that that the fate of ``$p$-brane democracy''
is a ``$p$-brane aristocracy'', in which neither strings 
nor $p$-branes are fundamental, but composite. 
In the following sections we will review dynamical theories
for possible constituents of strings, and suggest 
generalizations to $p$-branes.

\section{String-Bit Models}
\setcounter{equation}{0}
The string-bit models \cite{thornmosc,thornrpa,bergmantbits,
bergmantuniv}
are Galilean invariant matrix field theories in
$(D-2)+1$ dimensions possessing a global
$U(N_c)$ symmetry. The bits are thus
non-relativistic particles transforming in the adjoint
representation of $U(N_c)$.
In the continuum limit one recovers
$D$-dimensional light-cone string theory by interpreting
the bit Hamiltonian as $p^-$, and the number of bits in 
the chain as $p^+$. The {\bf non-relativistic} nature of the 
string-bit 
model then ensures the {\bf relativistic} identity
\begin{equation}
2p^+p^- - p_{\perp}^2 = M^2
\end{equation} 
in the continuum string theory.
Of course formation and stability of closed chains depends
on the microscopic bit interactions. In fact, any attractive
2-body interaction which supports a bound state will allow chain
formation. Stability, on the other hand, is only guaranteed for
certain supersymmetric string-bit models \cite{bergmantbits}, 
but 
universality is expected \cite{bergmantuniv}. 
Splitting and joining 
interactions of strings are manifest in these models as 
$\O(1/N_c)$ corrections, leading to the identification of
the string coupling constant as
\begin{equation}
g \sim 1/N_c \; .
\end{equation}


\subsection{Bosonic Model}
Let us begin by reviewing the bosonic string-bit model of
\cite{thornmosc}. We second quantize the string-bits
by introducing a field $\phi(x)_\alpha^\beta$ and its hermitian
conjugate $\phi^\dagger(x)_\alpha^\beta$, satisfying the 
canonical
commutation relation
\begin{equation}
[\phi(x)_\alpha^\beta,\phi^\dagger(y)_\gamma^\delta]
  = \delta_\alpha^\delta \delta_\gamma^\beta \delta(x-y)\; .
\end{equation}
The singlet operators
\begin{equation}
A^\dagger(x_1,\ldots,x_N) = N_c^{-N/2}{\rm Tr}[\phi^\dagger(x_1)\cdots
    \phi^\dagger(x_N)]
\end{equation}
behave as creation operators for $N$-bit chains in the 
large $N_c$ limit\footnote{$A^\dagger$ and
$A$ satisfy the canonical commutation relation only in 
the limit $N_c\rightarrow\infty$.}.
Consider the following string-bit Hamiltonian
\begin{equation}
H = {1\over 2m}\int dx{\rm Tr}|\nabla\phi|^2
  + {T_0^2\over 2mN_c}\int dxdy V(x-y){\rm Tr}[\phi^\dagger(x)
    \phi^\dagger(y)\phi(y)\phi(x)] \; .
\label{secondham}
\end{equation}
Acting on the closed chain state 
\begin{equation}
\ket{\psi} = \int dx_1\cdots dx_N
             A^\dagger(x_1,..,x_N)\ket{0}\psi_N(x_1,..,x_N)
\end{equation}
it gives
\begin{eqnarray}
H\ket{\psi} &=& \int dx_1\cdots dx_N \biggl\{
  A^\dagger(x_1,..,x_N)\ket{0} \sum_{k=1}^N
  \Big[-\nabla_k^2/2m + {T_0^2\over 2m}V(x_{k+1}-x_k)\Big]
  \psi_N(x_1,..,x_N) \nonumber\\
 &+&{T_0^2\over 2mN_c}\sum_{k=1}^N\sum_{l\neq k,k+1}
   A^\dagger(x_{k+1},..,x_{l-1})
   A^\dagger(x_l,..,x_k)\ket{0} 
   V(x_k-x_l)\psi_N(x_1,..,x_N)\biggr\} \; .
\label{hamonchain}
\end{eqnarray}
In the limit $N_c\rightarrow\infty$ the second term
drops out, and one can associate a first-quantized
Hamiltonian to an $N$-bit chain :
\begin{equation}
H_N = {1\over 2m}\sum_{k=1}^N
  \Big[p_k^2 + T_0^2V(x_{k+1}-x_k)\Big]\; .
\label{firstham}
\end{equation}
The above describes the dynamics of a bare closed chain,
provided of course that the nearest-neighbor potential 
is strong enough to bind. 
For the simple case in which the potential is harmonic, 
\begin{equation}
V(x) = x^2 \; ,
\label{harmonic}
\end{equation}
this system is exactly soluble by Fourier transforming.
The excitation energies of the collective modes of the chain 
are given by
\begin{equation}
E_n = {2T_0\over m}\sin{n\pi\over N} \;,
\end{equation}
where $n$ is the mode number.
In the continuum limit given by 
\begin{equation}
m\rightarrow 0 \;,\; N\rightarrow\infty
\;,\; mN={\rm fixed}
\end{equation}
the Hamiltonian becomes
\begin{equation}
 H_N \rightarrow {1\over 2T_0}\int_0^{mN/T_0}
 d\sigma \Big[{\cal P}(\sigma)^2 
 + T_0^2 x^\prime(\sigma)^2\Big] \; ,
\label{lcstring}
\end{equation}
and the excitation energies for the finite $n$ modes become
\begin{equation}
 E_n \rightarrow {2\pi nT_0\over mN}\; .
\label{contenergy}
\end{equation}
The finite $(N-n)$ modes have energies given by 
(\ref{contenergy}) with $n$ replaced by $(N-n)$. All
other modes have an infinite excitation energy, and 
therefore decouple in the continuum limit. 
The Hamiltonian (\ref{lcstring}) corresponds
precisely to the light-cone Hamiltonian ($p^-$)
of the bosonic string, if one identifies the
longitudinal momentum of the string as
\begin{eqnarray}
p^+ &=& mN\; .
\end{eqnarray}
In the continuum limit this is a continuous variable,
so it defines a new coordinate $x^-$ as its
canonical conjugate. From the point of view of string
theory this is the longitudinal coordinate of the string.
The string-bit model with $N_c\rightarrow\infty$ and a
harmonic potential is thus a straightforward discretization
of free light-cone bosonic string theory.

The appearance of an $\O(1/N_c)$ term 
in (\ref{hamonchain}), corresponding 
to the splitting of a single chain into two chains,
suggests an alternative point of view in which string-bits
are fundamental, and interacting (light-cone) string theory 
arises as a low energy effective theory. From 
this point of view the mass of string-bits ($m$) is
non-zero, and one studies finite size ($N$) chains.
At characteristic energies $\ll T_0/m$ only modes with
$n\ll N$ are probed, so the effective spectrum is stringy,
as in
(\ref{contenergy}). In this limit the $\O(1/N_c)$ chain
interactions become string interactions, so
the string coupling constant is identified with 
$1/N_c$.

This simple example shows that it is possible in principle
to construct a composite picture of strings in the light-cone
formulation. There are however four possible drawbacks to this 
particular model :
\begin{enumerate}
\item{It is unstable against the decay of chains into smaller
      chains.}
\item{It has strong long range interactions between chains.}
\item{The microscopic string-bit interaction leading to
      chain formation is not unique.}
\item{It depends strongly on the light-cone formulation
      of string theory.}
\end{enumerate}
The first point can be understood from the continuum point of 
view as the tachyonic instability. In the discrete model it
is simply a consequence of energetics.
The ground state energy of a long closed chain is generically
given by
\begin{equation}
 E_{\rm G.S.} = (D-2)\sum_{n=1}^{N-1}E_n 
 = {D-2\over m}\left[aN + {b\over N}
       + O({1\over N^2})\right].
\label{gsenergy}
\end{equation}
The first term is clearly the same for a single chain of $N$ 
bits 
and two chains of $N_1$ and $N_2$ bits, with $N_1+N_2=N$. 
For long chains, the nature of the true ground state then 
depends on 
the second term. In the above model $b<0$, so
$E_{0,N}>E_{0,N_1}+E_{0,N_2}$, implying that the chain is 
unstable to
decay into two smaller chains. In the continuum limit
$b$ becomes the mass squared of the tachyon.

The second point follows from the fact that the chain splitting
term in (\ref{hamonchain}) is weighted by the microscopic 
potential
$V(x-y)$, which in the harmonic case diverges at infinite
separation. This makes a definition of the S-matrix problematic.
For a well defined S-matrix we need a short range interaction,
which is however still strong enough to bind.
In addition the low energy (long wavelength)
properties of the chains should still approximate continuous 
strings.
Namely, the low energy collective excitations should reproduce
the string modes in the continuum limit. 
The idea of universality suggests that the long-wavelength
properties of the discrete system are to some extent independent
of the short-distance physics. We therefore expect a large
universality class of microscopic interactions, that includes
the harmonic potential, to yield an approximate description 
of free strings at low energy.
In particular the interaction can be short range. Thorn has
explored the possibility of a zero-range interaction, and has
computed numerically the string tension arising from such a model 
\cite{thornrpa}. Universality is an appealing idea, but it remains
unproven in bosonic string-bit models.

The third point is made clear by adding the following interaction
term to the Hamiltonian (\ref{secondham}),
\begin{equation}
 {\lambda\over N_c}\int dxdy U(x-y)
 :\Tr[\phi^\dagger(x)\phi(x)\phi^\dagger(y)\phi(y)]: \; ,
\label{ambiguity}
\end{equation}
where $\lambda$ and $U(x)$ are arbitrary.
Unlike the original interaction, the annihilation operators
do not appear consecutively in the matrix product. Consequently,
the action of this operator on the closed chain state will only
produce $\O(1/N_c)$ terms, which will not contribute in the
$N_c\rightarrow\infty$ limit to the first-quantized Hamiltonian
(\ref{firstham}). As far as free closed chains are concerned 
then,
the microscopic string-bit interaction is determined only up
to such a term.
This term will contribute to chain interactions, so
one might hope to fix the ambiguity by comparing chain scattering
amplitudes in the continuum limit to known results in 
string theory.

The fourth point may be significant and may be irrelevant,
depending on your point of view. 
Since the world-sheet (Riemann surface) picture breaks down at 
high temperature, world-sheet coordinate invariance is 
at best a low energy principle. If it is, then any 
reformulation of string theory appropriate for describing
the high energy behavior should reduce to the gauge invariant
world-sheet formulation at low energy.
Constituents are not a good candidate since
a constituent picture does not arise naturally in 
gauges other than light-cone.
However, in a formulation that is free of unphysical states from 
the outset
there is no need for gauge (world-sheet coordinate)
invariance.
Of course, such a formulation would generally be much more
complicated, and not Poincar\'e invariant. In the string-bit
models both world-sheet coordinate invariance and Poincar\'e
invariance are abandoned in favor of $(D-2)+1$-dimensional 
Galilean invariance
and a global $U(N_c)$ symmetry, such that the physical 
(a.k.a. gauge invariant) properties of strings are 
reproduced at low energy.
In addition, Poincar\'e invariance is recovered as 
an effective low energy 
symmetry at the critical dimension.
The fact that the low energy dynamics of chains seem
to arise from a light-cone gauge fixed world-sheet theory
is purely accidental from this point of view, and is
more a property of two dimensional field theory than 
of strings. In fact we will suggest a composite formulation
of $p$-branes in the next section, whose low energy limit
gives the the physical (transverse) modes, but cannot
be consistently derived from a covariant formulation,
as the latter does not possess sufficient gauge symmetry.

\subsection{Supersymmetric Models}
It is well known that the stability issue gets resolved
in string theory by making it supersymmetric. The ground
state of the superstring (IIA, IIB, or Heterotic) is massless.
We follow a similar path for string-bits by
associating an additional degree of freedom, ``statistics'',
to the bits,
distinguishing bosonic bits from fermionic bits. 
This will give rise to ``statistics'' waves traveling on 
the long chains, in addition to the usual coordinate
``phonon'' waves. 
Supersymmetry is implemented by extending the
Galilean symmetry of the bosonic string-bit model 
by gradation to an $\N=1$ Super-Galilean symmetry 
\cite{bergmantbits,bergmantgal}. 
The supercharges $\Q^A,\R^{\dot A}$ transform as real 
$D-2$-component spinors under the
$SO(D-2)$ subgroup of the Galilei algebra, 
and satisfy the following anti-commutation relations
\begin{eqnarray}
 \{\Q^A,\Q^B\} = mN\delta^{AB}\; &,& \;
 \{\Q^A,\R^{\dot B}\} = {1\over 2}\balpha^{A\dot B}
 \cdot {\bf P}\; ,
    \nonumber\\
  \{\R^{\dot A},\R^{\dot B}\} &=& \delta^{{\dot A}{\dot B}}H/2 \; ,
\label{superalg}
\end{eqnarray}
where $\alpha^i$ are the transverse 
$D$-dimensional Dirac 
$\alpha$-matrices corresponding to the embedding
$SO(D-2)\times SO(1,1)\subset SO(D-1,1)$. Undotted indices
correspond to positive $SO(1,1)$ chirality and dotted 
indices correspond to negative $SO(1,1)$ chirality.
Consequently $\Q$ and $\R$ will combine into a single
supercharge transforming as a spinor of $SO(D-1,1)$
upon recovery of the longitudinal dimension $x^-$ in the 
continuum limit. 

Models were constructed in 1+1 \cite{bergmantuniv}, 
2+1 and 8+1 \cite{bergmantbits} dimensions, 
underlying $D=3,4$ and $10$ type IIB superstring. In the lower
dimensional cases extra degrees of freedom must be eventually
added to make the string critical\footnote{An implementation
of this idea was first suggested in \cite{gilesmt}.}.
In the 2+1 and 8+1 models
\cite{bergmantbits} the last relation in (\ref{superalg})
could not be implemented. There were additional terms not
proportional to $\delta^{{\dot A}{\dot B}}$. In the 1+1
dimensional model there are no indices, so the full superalgebra
closes by default. In this model the supercharges
are given by 
\begin{eqnarray}
 {{\cal Q}}&=& \sqrt{m\over 2}\int dx
  {{\rm Tr}}\big[e^{i\pi/4}\phi^\dagger(x)\psi(x) + {\rm h.c.}\big]
  \nonumber\\
 {{\cal R}}&=& -{1\over 2\sqrt{2m}}\int dx
  {{\rm Tr}}\big[e^{-i\pi/4}\phi^\dagger(x)\psi^\prime(x) + {\rm
h.c.}\big]\nonumber\\
 & & + {1\over2N_c\sqrt{2m}}\int dx dy W(y-x) \nonumber\\
 & &~~~~~~~~~~\times{{\rm Tr}}\big[e^{-i\pi/4}\phi^\dagger(x)\rho(y)\psi(x)
     + {\rm h.c.}\big]\; ,
\label{secondquant}
\end{eqnarray}
where
$\psi_\alpha^\beta$ is the fermionic annihilation operator,
$\rho^\beta_\alpha = [\phi^\dagger\phi
+\psi^\dagger\psi]^\beta_\alpha$, and $W(x-y)$ is a
quantum-mechanical superpotential.
The 1+1 dimensional superalgebra is given by:
\begin{equation}
  \{{{\cal Q}},{{\cal Q}}\} = mN \; , \;
  \{{{\cal Q}},{{\cal R}}\} = -P/2 \; , \;
  \{{{\cal R}},{{\cal R}}\} = H/2 \; .
\label{anticomm}
\end{equation}
The last equation can be taken as the {\bf definition}
of the Hamiltonian, which is given by
\begin{eqnarray}
H &=& {1\over 2m}\int dx\Tr\Big[|\nabla\phi|^2 
                          + |\nabla\psi|^2\Big] \nonumber\\
&+& {1\over 2mN_c}\int dxdy\biggl\{
  \Big[W^2(y-x)+W^\prime(y-x)\Big]\Tr\phi^\dagger(x)\rho(y)\phi(x)
  \nonumber\\
&& ~~~~~~~~~~~~~~~~~~+\Big[W^2(y-x)-W^\prime(y-x)\Big]
               \Tr\psi^\dagger(x)\rho(y)\psi(x)\nonumber\\
&& ~~~~~~~~~~~~~~~~~~+W^\prime(y-x)
  \Tr\Big[i\phi^\dagger(x)\phi^\dagger(y)\psi(y)\psi(x)
      +\phi^\dagger(x)\psi^\dagger(y)\phi(y)\phi(x) 
      + {\rm h.c.}\Big]
  \biggr\}\nonumber\\
&+& {1\over 2mN_c^2}\int dxdydz W(y-x)W(z-x):\Tr\Big[
    \phi^\dagger(x)\rho(z)\rho(y)\phi(x)
   +\psi^\dagger(x)\rho(z)\rho(y)\psi(x)\Big]:\nonumber\\
&+& {1\over 4mN_c^2}\int dxdydz W(y-x)W(y-z)\times\nonumber\\
&&~~~~~~\times :\Tr\bigg[
 \phi^\dagger(x)\Big[\phi(z)\psi^\dagger(z),\rho(y)\Big]\psi(x)
 -i\phi^\dagger(x)
   \Big[\psi(z)\phi^\dagger(z),\rho(y)\Big]\psi(x)
  +{\rm h.c.}\bigg]:\; ,
\end{eqnarray}
where the commutators in the last integral refer only to 
matrix ordering. The last two integrals represent
3-body interactions. These were absent in the bosonic
model (\ref{secondham}), but are required here by
supersymmetry. 
This model is invariant
under the entire supersymmetry since
\begin{equation}
[H,\Q] = [H,\R] = 0 \; .
\end{equation} 
Acting with this Hamiltonian on the supersymmetric chain state
\begin{eqnarray}
 \ket{\Psi} &=& \int dx_1d\theta_1\cdots dx_Nd\theta_N 
 \Psi(x_1,\theta_1\ldots ,x_N,\theta_N) \nonumber\\
 &\times &{{\rm Tr}}
  \Big([\phi^\dagger(x_1) + \psi^\dagger(x_1)\theta_1]
       \cdots[\phi^\dagger(x_N) + \psi^\dagger(x_N)\theta_N]
  \Big) \ket{0}
     \; ,
 \label{chain}
\end{eqnarray}
and taking the limit $N_c\rightarrow\infty$
yields the following first quantized Hamiltonian :
\begin{eqnarray}
 H_N &=& {1\over 2m}\sum_{k=1}^N\Big\{
   p_k^2 + W^2(x_{k+1}-x_k) \nonumber\\
 &+& W^\prime(x_{k+1}-x_k)\big[\theta_k\pi_{k}
                                 - \pi_k\theta_{k}\nonumber\\
 &+& \pi_{k+1}\theta_k - \theta_{k+1}\pi_k-i(\theta_k\theta_{k+1}
+\pi_k\pi_{k+1})\big] \Big\} \; ,
\label{susyfirstham}
\label{secondsusyham}
\end{eqnarray}
where $\pi_k\equiv\partial/\partial\theta_k$.
Note that $W(x)$ is not the usual superpotential of
supersymmetric quantum mechanics (SUSY QM) \cite{wittensusyqm}, for
which the potential would be given by
\begin{equation}
V(x) = W^2(x) \pm W^\prime(x) \; .
\label{susypotential}
\end{equation}
That superpotential was restricted to having an odd number
of nodes by virtue of the zero ground state energy. As can
be seen in our Hamiltonian (\ref{secondsusyham}), the 
identification (\ref{susypotential}) does not hold in general,
and it is not yet clear what, if any, restrictions are placed on
$W(x)$. For the special case in which $W(x)$ is an odd function
however, the two-bit ($N=2$) sector is equivalent to the
SUSY QM of \cite{wittensusyqm}, and (\ref{susypotential}) holds.
A comprehensive study of the two-bit sector will be carried
out elsewhere.
For now we restrict ourselves to the linear superpotential
\begin{equation}
W(x) = T_0x \; ,
\label{linear}
\end{equation}
which defines the supersymmetric extension of the harmonic
string-bit model.
In this extension the excitation energies of the collective
statistics modes are the same as the phonon modes,
\begin{equation}
E_n = {2T_0\over m}\sin{n\pi\over N} \; ,
\label{susyexcite}
\end{equation}
but they contribute to the
ground state energy with opposite sign, 
giving an exact cancelation,
\begin{equation}
E_{\rm G.S.} = 0 \; ,
\label{susygs}
\end{equation}
for any length chain. 
The exact vanishing of the ground state energy
guarantees stability at the {\bf discrete} level,
which is a much stronger result than the absence
of the tachyon in continuum superstring theories. The latter
requires only that $b=0$ in (\ref{gsenergy}), whereas the 
former
states that all the terms in (\ref{gsenergy}) vanish. This is
crucial if the bit picture is to provide a {\bf fundamental}
formulation of superstring theory.

Upon changing the fermionic variables to
\begin{eqnarray}
S_k &=& {1\over\sqrt 2}(\theta_k + \pi_k) \nonumber\\
\tilde{S}_k  &=& {i\over\sqrt 2}(\theta_k - \pi_k) \; ,
\end{eqnarray}
the continuum limit gives
\begin{equation}
 H_N \rightarrow {1\over 2T_0}\int_0^{p^+/T_0}
 d\sigma \Big[{\cal P}(\sigma)^2 + T_0^2 x^\prime(\sigma)^2
 -iT_0S(\sigma)S^\prime(\sigma) 
 +iT_0\tilde{S}(\sigma)\tilde{S}^\prime(\sigma)\Big] \; .
\end{equation}
This is precisely the light-cone Hamiltonian of
the three dimensional free type IIB superstring, if one
interprets $S$ and $\tilde{S}$ as the right and left
moving fermions respectively.
We therefore refer to this 
supersymmetric 
string-bit model as a type IIB superstring-bit model.
Note that the supersymmetry has doubled to the required 
$\N=2$ in the continuum limit. The original supercharges
of the superstring-bit model $\Q,\R$ correspond to a sum of the
right and left moving supercharges of the continuum
type IIB superstring theory. 
One still has to check the finite $N_c$ interaction
terms, to see if they agree with the known continuum
results, or if additional terms need to be added to this
model.

The recovery of superstringy physics at low energy 
was manifest for the supersymmetric harmonic model,
but is actually a result of the fact that the ground
state energy vanished and the excitation gap was finite. 
Universality suggests that 
low energy superstringy physics is   
expected to hold more generally for
any supersymmetric model satisfying these two properties.
On the other hand the supersymmetric harmonic
model still gives rise to long range interactions 
between chains, precluding a well defined S-matrix.
One would therefore like to find 
short-range interactions leading to well defined
S-matrices, within the universality class of the 
harmonic model.

For 1+1 dimensional superstring-bit models 
\cite{bergmantuniv}, we were
able to prove a restricted form of universality
in perturbation theory, and argued that the ``smallness''
of the zero-energy representation of the superalgebra
implies that it can be extended beyond perturbation theory.
It was shown that deforming the 
superpotential away from linearity 
\begin{equation}
W(x) = T_0x + \delta W(x)\; ,
\label{deform}
\end{equation}
where
\begin{equation}
|\delta W(x)| \ll T_0 |x| \;\; {\rm for}\; 
|x|<\sqrt{\alpha^\prime} \; ,
\end{equation}
does not change the results (\ref{susyexcite}), (\ref{susygs})
qualitatively, but only renormalizes the string tension,
\begin{equation}
 T_0\rightarrow T_0
   + \langle\delta W^\prime(x_2-x_1)\rangle \; ,
\label{newtension}
\end{equation}
where the expectation value is computed in the zeroth order
(super-harmonic) chain state.
\begin{figure*}[htb]
\epsfxsize=1.6in
\centerline{\epsffile{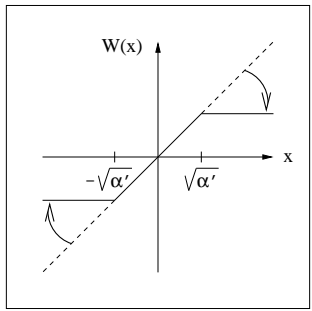}}
\caption{Superpotential deformation}
\end{figure*}
In fact with the deformation in Fig.(2)
we achieve the required short-range interaction, while still
keeping the stringy long-distance physics. Actually this deformation
is not quite enough to guarantee the clustering property required
for a well defined S-matrix. That is achieved by further replacing 
$\rho(y)$ with $\rho(y)-\sigma(y)$ in eq.~(\ref{secondquant}), where
$\sigma_\alpha^\beta=
:[\phi\phi^\dagger-\psi\psi^\dagger]_\alpha^\beta:$.
A successful implementation of these ideas for the higher
dimensional superstring-bit models, in particular ones
leading to critical superstrings, is under current study.

The superstring-bit model we presented above is most
likely incomplete, even in the simple $D=3$ case. 
In particular it seems to lack the
richness of structure of the superstring interactions
as given in \cite{greensb}. A detailed analysis of the 
perturbative chain interactions in this model will
be carried out elsewhere. 
Instead, we will try to understand the generic features
of such string-bit models under the extreme conditions
addressed in sections 2,3.

\subsection{Extreme Conditions}

At high temperature the Riemann surface description 
of string theory breaks down,
and a phase transition is expected to 
occur. It was shown that in such a transition
``holes'' must appear in the world-sheet, to support
a genus-0 free energy.
String-bit models offer an explicit realization of this
transition. It corresponds to
a dissociation transition, in which the chain breaks
into bits (Fig.(3)).
\begin{figure*}[htb]
\epsfxsize=4.5in
\centerline{\epsffile{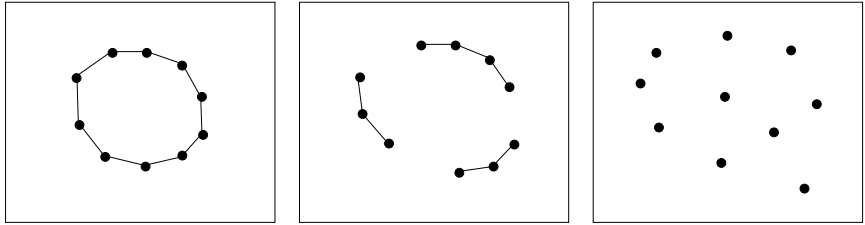}}
\caption{Dissociation transition}
\end{figure*}
The critical
temperature should be related to the energy scale at which 
the bonds between the bits break, in other words the binding
energy of two bits in a chain :
\begin{equation}
T_c \sim E_B \sim T_0/m \; .
\label{critical}
\end{equation}
Not surprisingly, this temperature is quite different than 
the Hagedorn temperature $T_H\sim \sqrt{T_0}$. Recall that
the Hagedorn transition was derived using perturbative
techniques. One can therefore trust the qualitative result
of the phase transition, but not necessarily the precise
value of the critical temperature.

The high temperature phase is a gas of non-relativistic
weakly interacting particles, with a free energy given by 
eq. (\ref{nrfree}) :
\begin{equation}
{F\over VT} \sim T^{(D-1)/2} \; .
\end{equation}
For a 2+1 dimensional string-bit model ($D=4$), 
appropriate for 
describing the microscopic building blocks of a 4-dimensional
string theory, this agrees with Atick and Witten's result
(\ref{stfree}).

The other extreme condition is a large longitudinal boost,
leading to a high density of string (Fig.(1)). From the string-bit 
point of view this corresponds to adding bits successively
to a chain (Fig.(4)).
The chain gets
longer because the length of a bond between nearest neighbor bits
is fixed by the microscopic bit interaction. 
For the harmonic
model the bond length is given by $\sqrt{\alpha^\prime}$,
so the length of an $N$-bit chain is $N\sqrt{\alpha^\prime}$. 
The amount of space
occupied by the chain 
can be estimated by the mean
square distance between two well separated bits in the chain,
\begin{equation}
\langle\Delta x^2\rangle = \langle(x_i - x_j)^2\rangle \; ,
\end{equation}
where $i-j\sim \O(N)$ for large $N$.
In the limit $\Nlarge$, appropriate for reproducing
the behavior of free string, we can ignore the non-nearest neighbor
interactions, and use the ground state of the bare chain 
Hamiltonian
$H_N$ for this purpose.
The result is given by
\begin{equation}
\langle \Delta x^2\rangle \sim \alpha^\prime\ln N \; ,
\end{equation}
in agreement with the free string result.
The density of string-bits then grows like
\begin{equation}
\rho \sim {N\over (\alpha^\prime\ln N)^{(D-2)/2}}\; .
\label{baredensity}
\end{equation}
\begin{figure*}[htb]
\epsfxsize=4.5in
\centerline{\epsffile{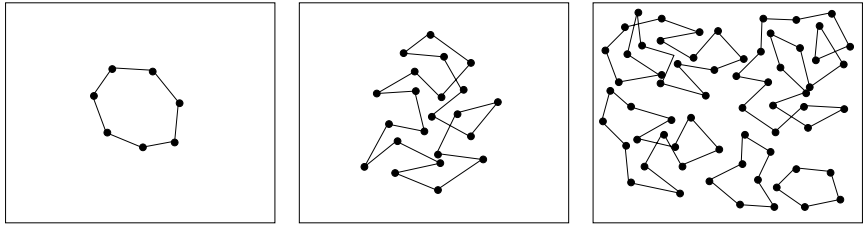}}
\caption{Adding bits to a chain}
\end{figure*}
As the density of string-bits grows, many non-nearest neighbor
bits will come close, and non-nearest neighbor interactions will
become important. Eventually the bare chain approximation used
to derive (\ref{baredensity}) will fail, just as the free string
approximation did in section 3.
Chains will join and split freely, resulting in an
ensemble of chains of varying lengths. 
The correction to eq.~(\ref{baredensity}), and with it the 
high density behavior of the chain, will depend crucially
on the nature of the non-nearest neighbor interactions.
In particular, an attractive non-nearest neighbor interaction
will tend to make the chain more dense, whereas a repulsive
one will tend to spread it out. A hard core repulsion 
(``elbows'') between
non-nearest neighbors would 
imply a maximum density of bits, $\rho_{\rm max}$,
depending on the size of the core.

In the bosonic model given by eq.~(\ref{secondham}) with $V(x)$ an 
attractive
potential, the non-nearest neighbor interaction
is $\sim V(x)/N_c^2$, and therefore also attractive. 
By adding to this model the term in 
eq.~(\ref{ambiguity}) with the same potential and 
$\lambda=-T_0/2m$,
one can make the non-nearest neighbor
interaction repulsive $\sim -V(x)/N_c^2$, while still maintaining 
an attractive nearest-neighbor interaction. 
Since $V(x)$ must be short range, the non-nearest neighbor
repulsion will have a finite length scale associated
with it, even though it is not necessarily hard core.
Indeed, the superpotential considered at the end
of section 5.2 will yield a non-nearest neighbor
interaction range of $\sim\sqrt{\alpha^\prime}$,
and a strength of $\sim T_0/(mN_c^2)$, coming from 
virtual chain corrections to the bare chain.
It is yet to be determined if this soft-core
repulsion gives
$\rho_{\rm max}\sim N_c^2(\alpha^\prime)^{(2-D)/2}$,
as suggested by Susskind.

Another possible origin of a maximum density of string-bits
is a hard core repulsion in the potential $V(x)$
itself. An example of
a superpotential which would achieve this is sketched in
Fig.(5). We do not yet know however if the universality
arguments made in \cite{bergmantuniv} can be extended to this 
superpotential.
\begin{figure*}[htb]
\epsfxsize=1.6in
\centerline{\epsffile{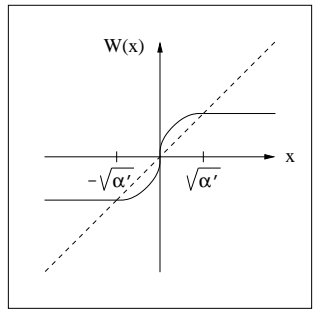}}
\caption{Hard core superpotential}
\end{figure*}

\section{p-Branes From Bits?}
\setcounter{equation}{0}
Finally we would like to suggest a generalization of the
string-bit models to incorporate other extended objects,
whose low energy limits are $p$-branes. For concreteness
we concentrate on bosonic membranes. We leave 
the extension to higher
$p$-branes and supersymmetry for future work.

String-bits could be thought of as particles with
a pair of ``legs'' (Fig.(6a)), 
that could be attached to other bits
consecutively to form a chain. The generalization
to membrane-bits then consists of adding more ``legs''
pairwise.
Closed membranes will be replaced with compact two-dimensional 
lattices of bits. The types of lattices one can form
will depend on how many ``legs'' the bits have, but
the continuum limit should be independent of this.
Let us assume membrane-bits to have two pairs of 
legs (Fig.6b)).
\begin{figure*}[htb]
\epsfxsize=3.5in
\centerline{\epsffile{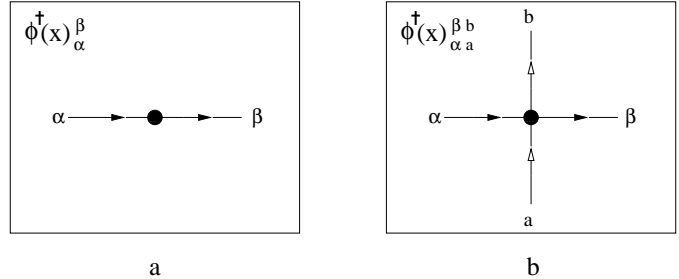}}
\caption{(a) String-bit (b) Membrane-bit}
\end{figure*}
This means that membrane-bit fields have two pairs of
indices,
$\phi(x)_{\alpha a}^{\beta b}$,
and the color group is enlarged to $U(N_c)\times U(N_c)$. 
Greek indices will be used for the first $U(N_c)$ factor,
and Latin indices for the second.
The bit-fields  
transform as the adjoint of each $U(N_c)$.
The canonical commutator is given by
\begin{equation}
[\phi(x)_{\alpha a}^{\beta b}, 
\phi^\dagger(y)_{\gamma c}^{\delta d}] = 
\delta_{\alpha}^{\delta}
\delta_{\gamma}^{\beta}
\delta_{a}^{d}
\delta_{c}^{b}
\delta(x-y) \; .
\end{equation}
The two $U(N_c)$ factors are distinct, and therefore so 
are the two sets of indices $(\alpha,\beta),
(a,b)$. Consequently indices belonging
to different $U(N_c)$ factors cannot be contracted.
The generalization of the free Hamiltonian is straightforward :
\begin{equation}
H_0 = {1\over 2m}\int dx 
\nabla\phi^\dagger(x)_{\alpha a}^{\beta b}
\cdot
\nabla\phi(x)^{\alpha a}_{\beta b}\; .
\label{memsecondham}
\end{equation}

Unlike string-bits, which could only form closed chains, 
membrane-bits can potentially form a rich variety of singlet
structures, with the continuum properties of membranes, strings,
or their combinations.
The building blocks are tiles composed of a finite number of
membrane-bits. The two possible pieces of string 
are shown in Fig.(7a), and possible pieces of membrane
are shown in Fig.(7b). Fig.(7c) depicts tiles
that can be sections of mixed structures.
\begin{figure*}[htb]
\epsfxsize=5in
\centerline{\epsffile{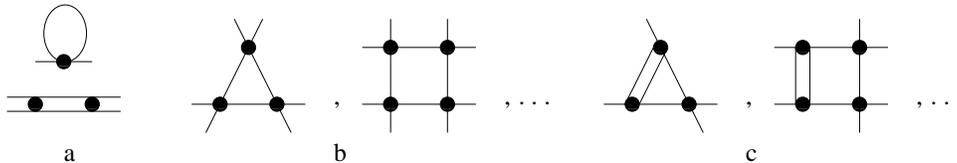}}
\caption{(a) String segments
(b) Membrane tiles (c) Pieces of mixed structures}
\end{figure*}
Closed membranes are formed by fitting together the
tiles in Fig.(7b) into compact (but not necessarily regular) lattices.
This will in general require using more than one kind of 
tile.
For examples of compact lattices and their continuum limits
see Fig.(8). The figures represent ways of tracing
over a product of tensor creation operators, and not
the precise shape of the physical structure, since the
bonds must all have equal length. The topology is however
the same.
\begin{figure*}[htb]
\epsfxsize=4.5in
\centerline{\epsffile{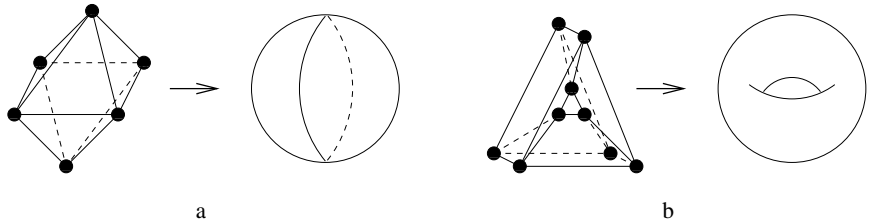}}
\caption{(a) Spherical membrane (b) Toroidal membrane}
\end{figure*}
Closed strings (chains) are formed by attaching either
of the string segments of Fig.(7a) to each
other (Fig.(9a)). 
Other stringy objects can be
formed by combining the two kinds of segment
(Fig.(9b,c)).
Note that the continuum limit of Fig.(9b) looks like
an open string whose ends are attached to a closed
string, and the continuum limit of Fig.(9c)
looks like an open string whose ends are attached
to two different closed strings. 
Such configurations are familiar in the D-brane 
approach to 1-branes, which play the role of the closed 
string \cite{polchinskidbrane}.
A generic membrane-bit 
model would include these configurations, and would 
therefore encompass the D-brane idea.
\begin{figure*}[htb]
\epsfxsize=4.5in
\centerline{\epsffile{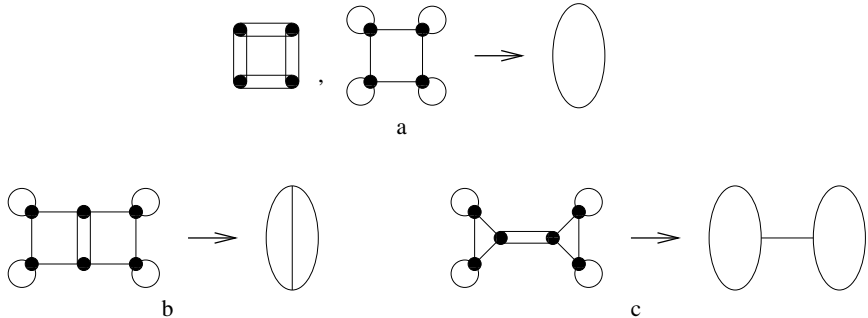}}
\caption{Stringy objects}
\end{figure*}

A priori all of these structures are just mathematical (singlet) 
entities.
Given a particular interaction term in the membrane-bit
Hamiltonian, some of them will be actual physical structures.
For these, the inter-bit connections correspond to physical 
bonds. The interaction should be general enough to 
allow the formation of both stringy and membrane-like 
discrete structures, but not necessarily all possible
discrete structures. 
As in the bosonic string-bit model, we will consider
only two-body interactions. Unlike the string-bit model
however, there are many ways to form a connected  singlet 
two-body interaction term. 
We do not intend to present an exhaustive analysis of two-body
membrane-bit interactions. Instead we will give a taste 
of the richness of membrane-bit models by considering
only the following three interactions:
\begin{eqnarray}
H_1 &=& {1\over N_c^2}\int dxdy V_1(x-y)
\phi^\dagger(x)_{\alpha a}^{\beta b}
\phi^\dagger(y)_{\beta b}^{\gamma c}
\phi(y)_{\gamma c}^{\delta d}
\phi(x)_{\delta d}^{\alpha a} \nonumber\\
H_2 &=& {1\over N_c}\int dxdy V_2(x-y)
\phi^\dagger(x)_{\alpha a}^{\beta b}
\phi^\dagger(y)_{\beta c}^{\delta d}
\phi(y)_{\delta d}^{\epsilon c}
\phi(x)_{\epsilon b}^{\alpha a}\nonumber\\
H_3 &=& {1\over N_c}\int dxdy V_3(x-y)
\phi^\dagger(x)_{\alpha a}^{\beta b}
\phi^\dagger(y)_{\gamma b}^{\delta d}
\phi(y)_{\delta d}^{\gamma c}
\phi(x)_{\beta c}^{\alpha a} \; .
\label{meminteractions}
\end{eqnarray}
These interactions are represented graphically 
in Fig.(10). Note that the unfilled bits represent
annihilation operators (removal of bits), whereas
the filled bits represent creation operators.
\begin{figure*}[htb]
\epsfxsize=4.5in
\centerline{\epsffile{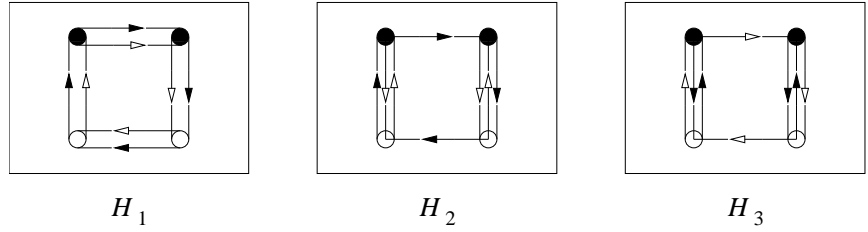}}
\caption{Examples of two-body membrane-bit interactions}
\end{figure*}
As in the string-bit models
these interaction terms 
should act invariantly on 
singlet structures to lowest order in $1/N_c$,
and give rise to a nearest-neighbor interaction pattern.
Since the interactions we are considering act on two
bits at a time, it is sufficient to determine their
action on connected 2-bit states, which can be embedded
in larger singlet structures (chains or lattices).
Of course just as there are many possible interaction terms,
there are many possible 2-bit states.
The possible 2-bit states are shown in Fig.(11), 
\begin{figure*}[htb]
\epsfxsize=4in
\centerline{\epsffile{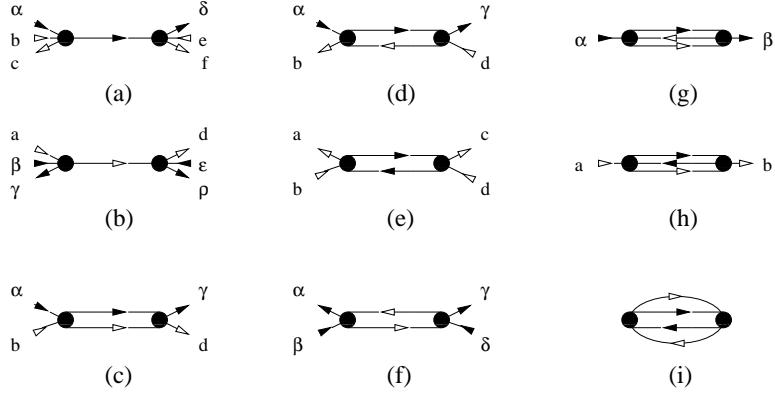}}
\caption{Connected 2-bit states}
\end{figure*}
and are respectively given by
\begin{eqnarray}
 ({\rm a}) \;
 \phi^\dagger(x)_{\alpha b}^{\lambda c}
 \phi^\dagger(y)_{\lambda e}^{\delta f}\ket{0}\;
 &\;\;
 ({\rm d}) \;
 \phi^\dagger(x)_{\alpha l}^{\lambda b}
 \phi^\dagger(y)_{\lambda d}^{\gamma l}\ket{0}
 &\;\; 
 ({\rm g}) \;
 \phi^\dagger(x)_{\alpha t}^{\lambda l}
 \phi^\dagger(y)_{\lambda l}^{\beta t}\ket{0}\;
 \nonumber\\
 ({\rm b}) \;
 \phi^\dagger(x)_{\beta a}^{\gamma l}
 \phi^\dagger(y)_{\epsilon l}^{\rho d}\ket{0}
 &\;\;
 ({\rm e}) \;
 \phi^\dagger(x)_{\lambda b}^{\theta a}
 \phi^\dagger(y)_{\theta d}^{\lambda c}\ket{0}\;
 &\;\;
 ({\rm h}) \;
 \phi^\dagger(x)_{\lambda a}^{\theta t}
 \phi^\dagger(y)_{\theta t}^{\lambda b}\ket{0} 
 \nonumber\\
 ({\rm c}) \;
 \phi^\dagger(x)_{\alpha b}^{\lambda l}
 \phi^\dagger(y)_{\lambda l}^{\gamma d}\ket{0}\;
 &\;\;
 ({\rm f}) \;
 \phi^\dagger(x)_{\beta l}^{\alpha t}
 \phi^\dagger(y)_{\delta t}^{\gamma l}\ket{0}
 &\;\;
 ({\rm i}) \;
 \phi^\dagger(x)_{\alpha a}^{\beta b}
 \phi^\dagger(y)_{\beta b}^{\alpha a}\ket{0}
\label{nnsections}
\end{eqnarray}
Exchanging $x$ and $y$ yields new states, except for
(e),(f) and (i), and we will denote them by the corresponding
letter with a prime.
We now act with each of the Hamiltonians in
(\ref{meminteractions}) on each of these states,
keeping track only of the general structure, not of the
precise indices. This will yield two terms for each state,
corresponding to the two ways of contracting the two
annihilation operators of the interactions with the two
creation operators of the states.
Every bond overlap will yield a factor of $N_c$.
The results are summarized in the following table :

\begin{center}
\begin{tabular}{|c||c|c|c|} \hline
 {\mbox{}} & $H_1$ & $H_2$ & $H_3$ \\ \hline\hline
(a) & $N_c^{-1}$(c)+$N_c^{-2}$(h') 
    & (a)+$N_c^{-1}$(e) 
    & $N_c^{-1}$[(c)+(d)] \\ \hline
(b) & $N_c^{-1}$(c)+$N_c^{-2}$(g')
    & $N_c^{-1}$[(c)+(d')]
    & (b)+$N_c^{-1}$(f) \\ \hline
(c) & (c)+$N_c^{-2}$(i) 
    & (c)+$N_c^{-1}$(h) 
    & (c)+$N_c^{-1}$(g') \\ \hline
(d) & $N_c^{-1}$[(g)+(h')]
    & (d)+$N_c^{-1}$(h') 
    & $N_c^{-1}$(g)+(d) \\ \hline
(e) & $N_c^{-1}$[(h)+(h')] 
    & $2$(e) 
    & $N_c^{-1}$[(h)+(h')] \\ \hline
(f) & $N_c^{-1}$[(g)+(g')] 
    & $N_c^{-1}$[(g)+(g')] 
    & $2$(f) \\ \hline
(g) & (g)+$N_c^{-1}$(i) 
    & (g)+$N_c^{-1}$(i) 
    & $2$(g)\\ \hline 
(h) & (h)+$N_c^{-1}$(i) 
    & $2$(h) 
    & (h)+$N_c^{-1}$(i) \\ \hline
(i) & $2$(i) 
    & $2$(i) 
    & $2$(i) \\ \hline
\end{tabular}
\end{center}
In the limit $\Nlarge$, the states left invariant 
by $H_1,H_2$ and $H_3$ are respectively
$\{({\rm c}),({\rm g}),({\rm h}),({\rm i})\}$,
$\{({\rm a}),({\rm c}),({\rm d}),({\rm e}),({\rm g}),({\rm h}),
({\rm i})\}$ and $\{({\rm b}),({\rm c}),({\rm f}),({\rm g}),({\rm h}),
({\rm i})\}$. When the interaction $H_i$ ($i=1,2,3$) acts on a large
singlet structure in the limit $\Nlarge$, one gets a 
nearest-neighbor interaction $V_i$ for every membrane-bit pair
belonging to the corresponding set.
With more than one of the interaction terms present, some 
of the membrane-bit pairs will acquire a sum of interactions.
For example, if the membrane-bit model had the following interaction
\begin{equation}
H = \lambda_1H_1 + \lambda_2H_2 + \lambda_3H_3 \; ,
\end{equation}
the membrane-bit pair (c) would interact through 
$\lambda_1V_1+\lambda_2V_2+\lambda_3V_3$,
whereas the pair (a) would only interact through
$\lambda_2V_2$, and the pair (b) would only interact
through $\lambda_3V_3$.

Pure membranic structures contain only single bonds,
and therefore contain only the pairs (a) and (b). 
Consequently
if $\lambda_2=\lambda_3=0$, the model would not support 
pure membrane formation, since (a) and (b) are not in 
the invariant set of $H_1$. 
Closed chains on the 
other hand can be single-bonded or double-bonded (Fig.(9a)).
For these values of the parameters only the
double-bonded variety would be supported.
At generic values of the parameters,
both strings and membranes, as well as mixed structures, 
are possible. A reasonable (but not necessary) assumption
is that the bonds all carry the same interaction,
$V_1(x)/2=V_2(x)=V_3(x)=V(x)$. The parameter space
spanned by $\lambda_1,\lambda_2,\lambda_3$ then serves
as a kind of ``moduli space''. At a generic point all
structures can exist, but there are regions where only 
some of the structures will form.
Of course there are many more possible membrane-bit interaction
terms, and therefore many more parameters in this ``moduli space''. 

The generalization to higher dimensional $p$-branes consists
of adding more pairs of legs to the bits. A $p$-brane bit will
have $2p$ legs. Such bits would be able to form not only
$p$-branes, but by utilizing self-contractions, 
double-contractions,
triple-contractions, etc., would also be able to form all the
lower branes, as well as various mixed structures.

The $0$-brane is somewhat puzzling. The only candidate seems
to be a single bit with all its legs self-contracted.
But as a single bit, this object should disappear in 
the continuum limit.

\section{Discussion}
\setcounter{equation}{0}
All of the evidence presented here, together with further
investigations of string dualities, suggest that the true
formulation of string theory, if it exists, is radically
different from the world-sheet formulation. 
We have considered
the possibility that string theory is in fact a low energy
effective theory of a more fundamental theory of constituents.
String-bit models are an appealing idea because they provide
a microscopic origin for both the curious high temperature 
behavior and the high longitudinal boost behavior
of string theory. They predict a dissociation phase transition
at $T_c\sim T_0/m$, and can result in different high $p^+$ 
behaviors depending on the details of the microscopic bit 
interactions. Agreement with the black hole entropy formula
seems to imply that there is a maximum density of bits,
but it is not yet clear which string-bit model will achieve
this. 

The main drawback of string-bit models is that there
are so many of them. There is a large universality class 
of models that leads to the same free string limit. Comparing
with known string interactions may fix some of the ambiguity.
In fact it was shown in the supersymmetric case that a rather
specific form of bit interaction was required in order to prevent
long range interactions between separated chains.
However, it is not clear yet whether there is a unique 
superstring-bit model.

It is also not clear how the role of the dilaton as the
string coupling constant is realized in string-bit models,
given that $g\sim 1/N_c$. 

Finally, we have seen that membranes and other extended
objects could in principle be formed from constituents
as well. Given bits that have $2p$ legs, and sufficient
two-body interactions, one can form
extended structures of dimension $\le p$. The various two-body
interactions span a ``moduli space'' containing degenerate
regions, where some of the structures are not supported. 
This bears close resemblance to the idea of ``$p$-brane 
democracy'' that asserts that in the fundamental formulation
all $p$-branes are equally fundamental. It is only when we look
at a special region of moduli space that the string appears
to be more fundamental.

\vskip 1cm

\centerline{\bf Acknowledgments}

\noindent This work was supported in part by the Department 
of Energy under grant DE-FG05-86ER-40272 and by the Institute
for Fundamental Theory.
I wish to thank Charles Thorn and Larus Thorlacius for 
useful discussions.

\end{document}